# Using Prediction Markets to Incentivize and Measure Collective Knowledge Production

THOMAS MAILLART, Chair of Entrepreneurial Risks, ETH Zurich, Switzerland
DIDIER SORNETTE, Chair of Entrepreneurial Risks, ETH Zurich, Switzerland.

One of the toughest challenges to ensure the emergence of collective intelligence and knowledge production is to rightly set individual incentives towards the achievement of a common goal [Pickard et al. 2011]. Unfortunately, most attempts to establish and maintain long-term collective action are confronted with diverging individual incentives [Hardin 1968], and organizational issues [Ostrom 1990]. On the Internet, collective action is best embodied by open source software (OSS) and other non-software knowledge production projects, like Wikipedia. Our current knowledge of OSS communities reveals how they achieve knowledge production in a seamless self-organized way called *peer-production*, which relies on two fundamental organizational rules : (i) task self-selection and (ii) peer-review [Benkler 2002]. Beyond these rules, the success or the failure of OSS projects is most often bound to people's capabilities to build true communities, through institutions that help solve the tensions existing between individual incentives and the collective interest. Building and adapting institutions for open collaboration is often a hard and painful enterprise, which can put at stake the very survival of an OSS project [O'Mahony and Ferraro 2007]. Here we present a mechanism design, based on a *Wiki*-like collaboration platform coupled to a prediction market, which aligns individual incentives with the goals of collective knowledge *peer-production*, and does not require further governance mechanism. We have implemented and tested four instances of this system in the context of higher education for the courses Entrepreneurial Risks (ETH Zurich; Course No 351-0564-00L; instructor Didier Sornette) in 2011, 2012, 2013, and Environmental Entrepreneurship (EPFL; instructor Marc Vogt) in 2013. Complementing business management experiences [Hoyt and Rao 2006; Benbya and Van Alstyne 2011], this system explicitly formalizes the rules of peer-production. Most prediction markets are designed to organize information rather than to engage individual knowledge production as a collective good. Our implementation helps value not only information, but also individual and collective behaviours, which are at the very roots of collective intelligence.

## 1. COUPLING COLLECTIVE KNOWLEDGE PRODUCTION WITH A PREDICTION MARKET

Students start their contributions by posting ideas, which in turn further develop into projects, according to the specific guidelines provided by the instructor.[1] A virtual currency called Entrepreneurial Risks Dollar (ER$) is issued and students are initially endowed with ER$ 10'000. Students are completely free to post as many new project ideas as they want. They can also contribute without restriction to any existing project. For their contributions, they are rewarded in the following way: each 10 words (or 55 bytes) contributed are worth ER$ 100 of shares of the contributed project. For instance, if a project stock value is ER$ 100, for every 10 words contributed one share is attributed. If the project value is ER$ 200, for every ten words contributed half a share is attributed. On the contrary, if the

---

[1]Technically, a project corresponds to a page on a *Wiki*. All changes made by students to any page (i.e. any project) are parsimoniously tracked over the course of the semester. Guidelines can encompass a specific topic or problem to solve, as well as instructions on the required quality of the work.

project value is ER$50, then for every ten words contributed, two shares are attributed, and so on. For each new project, five shares (worth ER$ 500) are attributed to the original contributor. The shares are subsequently traded and valued on the prediction market as a form of peer-review. To ensure equal access to information, all contribution and prediction market records are accessible to everyone, in the most transparent way. As they contribute to projects and trade on the prediction market, students build a portfolio of stock holdings. The evolution of the portfolio provides a real-time indication of their performance over the semester. To avoid a "beauty contest" problem [Keynes 1936], before the class starts students are informed that their final grade will correspond to the value of their stock holdings for each project weighted by their *ex-post* value, i.e. the final project score provided by the instructor. In other words, while students can freely trade over the semester, their portfolio value is a reliable predictor of their final grade, only if they are able to accurately anticipate the final *ex-post* judgement by the instructor, and trade accordingly. In a nutshell, students should not trade according to their expectations, but rather those of the instructor, which provides a kind of fundamental value anchor similar to more standard education grading systems. This should also avoid the emergence of bubbles, i.e. transient speculative behaviors that may be disconnected from the goal of our design.

## 2. RESULTS

The overall goal of coupling collective knowledge production with a prediction market is to recreate a controlled environment of collective intelligence and knowledge production. As expected, we find that our mechanism design is efficient to foster diversification of contributions and trading : students are incentivized to sneak into projects – even if they do not make any major contribution, and do it overall quite well (see Figure 1 panel A for a matrix representation of contributions across projects). We then find that despite a low level of liquidity and sometimes a lack of counterparts for trading, the prediction market does rather well in predicting the final *ex-post* scores. Finally, while students are given metrics to readily measure their *ex-ante* score over the semester, we find, surprisingly, that some students contribute orders of magnitude more than the average. This result is at odds with the rational utility maximizer behavior we could expect from a mechanism design, which sets only extrinsic motivations as incentives (i.e getting a good grade in this case). This result is however compatible with the complicated mixture of extrinsic and intrinsic motivations, driving open source software development [von Krogh et al. 2012]. An anonymous survey conducted after the end of the class in 2012, as well as informal discussions with students after they got their grades, confirms that motives go sometimes far beyond just performing on the prediction market. This very encouraging result suggests that the prediction market itself does not undermine intrinsic motivation, which is known to be critical for the success of open source projects [von Krogh et al. 2012]. We also see that highly contributing students have systematically more opportunities to compound on their work : as time passes, a clear pattern of positive return of scale on score versus contributions develops (see Figure 1 panel B). Since all actions are parsimoniously recorded and timestamped, we can trace back to the very detailed origins of these stylized facts.

## 3. UNDERSTANDING AND MANAGING SELF-ORGANIZED COLLECTIVE INTELLIGENCE

While it is still at an early stage, this education project provides surprising insights on the emergence of collective intelligence. We discuss some aspects and limitations relevant to the understanding of underlying mechanisms, as well as for management and organization design. In the process of filtering and organizing knowledge, peer-review plays a central role. We initially expected that students would carefully review projects. It turns out that this is nearly impossible to do so, because projects are in continuous, and often fast-paced, evolution. Thorough review is certainly not worth the effort, and thus students engage instead into "quick and dirty" judgements on projects. Nevertheless, the prediction market does overall well. Why is it so ? On the one hand, students have multiple occasions to judge a project, and they can do it with different levels of confidence by putting more or less shares at stake. Therefore one or a few mistaken judgements – over several tens or hundreds market orders – are not critical. Moreover, in any case, the judgement can only be an approximation the "true" value of a project. On the other hand, prediction market transactions always involve at least two students who have reached an agreement on the value of a project. This gives obviously much more strength to the judgement. We wonder, however, how a project, addressing a truly technical and complicated problem, could be evaluated efficiently by this sort of "wisdom of crowd". It probably heavily depends on the skill sets of all participants. Another insight for management is the powerful creative destruction process at work : unsuccessful projects are quickly abandoned, and their value drops rapidly, because there is only a limited amount of money in circulation that must be optimally invested on the market. There is however a balancing effect, which helps "saving" potentially good, but undervalued projects : if the stock market price drops, the amount of shares received per contribution unit increases, thus creating incentives to contribute. On the contrary, if a project becomes highly valued, the amount of shares distributed per contribution unit vanishes, thus reducing incentives to further modify the project. This helps ensuring stability of more mature projects. Finally, we found that unfortunately this form of collective knowledge production is little suited for the achievement of self-contained final reports. We believe that fixing the editorial line requires intense work by a single contributor, which is disproportionate in comparison to average contributions we generally expect from students, and thus deters any contributor to take on the task. This makes in turn final project grading more difficult for the instructor.

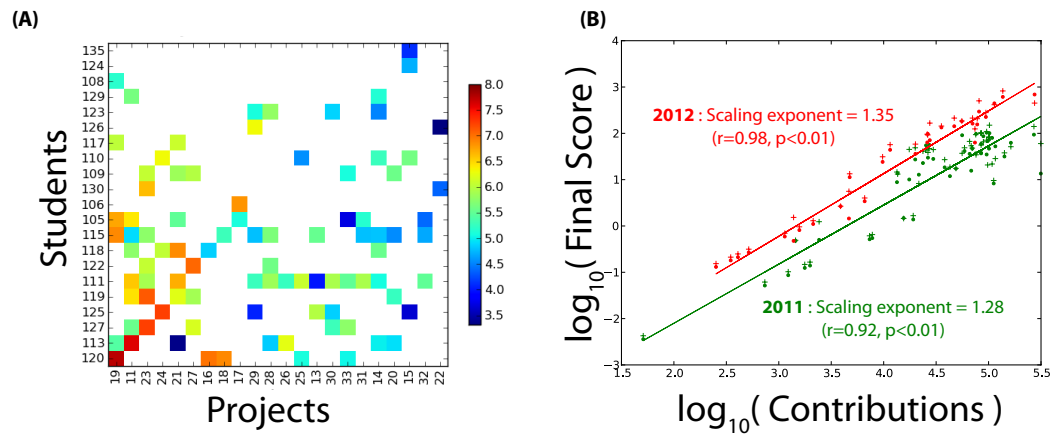

Fig. 1. **A.** Amount of contributions in bytes (logarithmic scale) by students to projects, sorted by decreasing order of contribution on both axes, with higher contribution on the bottom-left of the matrix. **B.** Scaling function of final score as a function of contributions for years 2011 and 2012. The linear fit in double logarithmic scales with slope larger than 1, shows the positive return of scale of final score as a function of contributions. The dots (resp. crosses) correspond to the *ex-ante* (resp. *ex-post*) final score. The fits presented here are made on the *ex-ante* finale score.

We have presented a mechanism design that helps replicate typical instances of collective intelligence and knowledge production found on the Internet, in a controlled environment. While its experimentation and testing are still at an early stage, we anticipate that this mechanism design will help further uncover detailed insights on underlying emerging and sustainable collective intelligence mechanisms.

Acknowledgements

The authors acknowledge support from *Innovedum* (Fostering teaching innovation projects at ETH Zurich, grant Innovedum 2012-2013), from Qunzhi Zhang, Andronikakis Alexandros and Guillaume Schaer for their programming input, Dorsa Sanagdol for her support in community management, and useful feed-back from all Chair of Entrepreneurial Risks team members, since 2011. T. Maillart is now affiliated at the UC Berkeley School of Information, and acknowledges support from the Swiss National Science Foundation (Grant No. PA00P2_145368).